\documentclass[12pt]{JHEP3}

\usepackage{amssymb}
\usepackage{graphics}
\usepackage{epsfig}
\usepackage{graphicx}
\usepackage{subfigure}
\usepackage{bm}

\def\beq{\begin{equation}}
\def\eeq{\end{equation}}
\def\baq{\begin{eqnarray}}
\def\eaq{\end{eqnarray}}

\def\d{{\rm d}}

\def\bea{\begin{eqnarray}}
\def\eea{\end{eqnarray}}
\def\be{\begin{equation}}
\def\ee{\end{equation}}

\title{Higgs Dynamics during Inflation}

\author{Kari Enqvist, Tuukka Meriniemi and Sami Nurmi
\\
University of Helsinki and Helsinki Institute of Physics, P.O. Box
64, FI-00014, Helsinki, Finland
}

\abstract {We investigate inflationary Higgs dynamics and
constraints on the Standard Model parameters assuming the Higgs
potential, computed to next-to-next leading order precision, is not
significantly affected by new physics. For a high inflationary scale
$H\sim 10^{14}$ GeV suggested by BICEP2, we show that the Higgs is a
light field subject to fluctuations which affect its dynamics in a
stochastic way. Starting from its inflationary value the Higgs must
be able to relax to the Standard Model vacuum well before the
electroweak scale.  We find that this is consistent with the high
inflationary scale only if the top mass $m_t$ is significantly below
the best fit value. The region within $2\sigma$ errors of the
measured $m_t$, the Higgs mass $m_h$ and the strong coupling
$\alpha_s$ and consistent with inflation covers approximately the
interval $m_t \lesssim 171.8\,{\rm GeV} + 0.538(m_h-125.5\,{\rm GeV} )$ with
$125.4\,{\rm GeV}\lesssim m_h\lesssim 126.3\,{\rm GeV}$. If the low
top mass region could be definitively ruled out, the observed high
inflationary scale alone, if confirmed, would seem to imply new
physics necessarily modifying the Standard Model Higgs potential
below the inflationary scale.}

\preprint{HIP-2014-07/TH}

\begin{document}

\section{Introduction}

With the confirmed discovery of the Standard Model (SM) Higgs boson
at LHC \cite{LHC}, it is now apposite to study in detail the evolution of the
Standard Model Higgs during inflation. Here we do not adopt any
particular inflationary model but merely assume that there is a
period of superluminal expansion with a very slowly changing Hubble
rate $H$. Intriguingly, as has been much discussed lately, the Higgs
field could be the inflation
\cite{Bezrukov:2007ep,Barvinsky:2008ia,Bezrukov:2008ut,Bezrukov:2010jz},
albeit at the expense of an abnormally large non-minimal coupling to
gravity. However, if confirmed, the detection of primordial
gravitational waves by BICEP2 determines the inflationary energy
scale to be $\rho^{1/4}\sim 10^{16}$ GeV at the horizon crossing of
the observable patch, together with a tensor-to-scalar ratio
$r=0.20{}^{+0.07}_{-0.05}$ \cite{Ade:2014xna} , which appears to be
at odds with Higgs inflation, see however
\cite{Hamada:2014iga,Masina:2014yga,Germani:2014hqa}. Here we do not consider this
or any other modified SM scenario but rather investigate SM Higgs
dynamics assuming its couplings are not significantly affected by
whatever the new physics is driving inflation.

The starting point for our analysis is the next-to-next to leading
order expression for the SM effective Higgs potential. As is well
known, at high Higgs field values the SM potential typically becomes
unstable as one moves beyond a critical field value $h  = h_c$ and
there is a local maximum located at $h_{\rm max}< h_c$. Both the
point of instability $h_c$ and $h_{\rm max}$ are very sensitive to
the SM parameter values as measured at the electroweak scale
\cite{Degrassi:2012ry,Chetyrkin:2012rz,Bezrukov:2012sa}; for the
best fit SM parameters $h_{c} \sim 10^{10}$ GeV. However, the
instability can be pushed up to $10^{16}$ GeV and beyond by lowering
the top mass value. Consistency of the setup of course requires the
Higgs potential to be stable at the inflationary scale implied by
BICEP2 \cite{Ade:2014xna}. In particular, as pointed out already 
in \cite{riotto} and later discussed in \cite{higgsinstability}, 
the inflationary
fluctuations should not push the Higgs field over the local maximum
$h_{\max}$ and into the false vacuum during the last $60$ e-folds of
inflation, corresponding to the observable universe. In other words:
we require that at the end of inflation we find ourselves in the
region of field space from where the SM vacuum can dynamically be
reached. Imposing this constraint we identify the region in the
space of the top mass, the Higgs mass, and strong coupling, where
the SM Higgs potential remains compatible with the measured
inflationary scale of $\rho^{1/4}\sim 10^{16}$ GeV.

Given the generic form of the Higgs potential the question then is:
how does the Higgs field evolve during inflation? The answer very
much depends on whether the Higgs is a light field or not, but also
on the initial field value. Our starting point is that well before
the electroweak symmetry breaking takes place, the Higgs field must
find itself far away from the instability and close to the
low-energy vacuum $h=\nu\simeq246$ GeV. If at the onset of
inflation, the Higgs is on the wrong side of the local maximum at
$h_{\rm max}$ , it must tunnel during inflation to the other side
unless the false vacuum is lifted by thermal corrections after the
end of inflation. Since the SM expression of the Higgs potential
much beyond $h_{\rm max}$ must be modified by unknown new physics,
we cannot assign a model-independent probability measure for such a
tunneling event. However, since tunneling rates depend exponentially
on the differences of the free energies, if tunneling from $h\gg
h_{\rm max}$ takes place, afterwards the most probable field value
is $h_{\rm max}$, the local maximum. Tunneling could take place any
time before the end of inflation, and of course, the initial field
value could also be  $h\ll h_{\rm max}$ simply by chance.

We are thus led to study the dynamics starting from arbitrary
initial values in the range $h \leqslant h_{\rm max}$.  We find that
the SM Higgs is either a light field to start with or becomes light
after at most a few e-folds, and its energy density is small
compared to the inflationary scale. As its contribution to the total
energy density is tiny, the Higgs condensate (zero mode) field
acquires nearly scale invariant fluctuations on superhorizon scales.
We find that quantum fluctuations dominate over the classical motion
close to the maximum $h_{\rm max}$ as well as in the asymptotic
regime $h\ll h_{\rm max}$. In the asymptotic quantum regime the mean
field fluctuates and is random walking while local perturbations are
also being generated. The typical values of the Higgs condensate
after the end of inflation, together with its fluctuations, are
directly determined by the inflationary scale. If the mechanism for
generating curvature perturbation is sensitive to the Higgs value,
for example through a modulation of the inflaton decay rate
\cite{Lyth:2005qk,Choi:2012cp}, the Higgs fluctuations could leave
an imprint in the primordial metric perturbations. In this case the
transition from classical Higgs dynamics to the quantum regime could
also generate characteristic features in the primordial
perturbations, provided the transition occurs when observable scales
are crossing the horizon.

The paper is organized as follows. In section \ref{sec:inf} we
review the form of the radiatively corrected SM Higgs potential and
derive consistency conditions for a high scale inflation with the SM
Higgs as a spectator field during inflation. In section
\ref{sec:dynamics} we present a detailed analysis of the dynamics of
the SM Higgs during inflation. Finally, in section
\ref{sec:discussion} we summarize the results and discuss possible
consequences of SM modifications.

\section{Standard Model Higgs and high scale inflation}\label{sec:inf}

For large field values $h\gg \nu \simeq 246$ GeV the radiatively corrected
effective potential of the Standard Model Higgs can be expressed in
the form
  \beq
  \label{Vh}
  V(h)=\frac{\lambda(h)}{4}h^4\ .
  \eeq
The running of $\lambda(h)$ has been computed explicitly to
next-to-next to leading order precision \cite{Degrassi:2012ry,
Chetyrkin:2012rz, Bezrukov:2012sa}. Throughout this work we will
refer to the effective potential evaluated in the ${\overline{MS}}$
renormalization scheme, and all the couplings and field values are
given within this scheme. As these are not directly physical, a
different choice of the renormalization scheme would give different
numbers to be associated with the same physical quantities.

The self coupling $\lambda(h)$ is determined from its
$\beta$-function
  \beq
  \beta_{\lambda}=\frac{\d \lambda}{\d{\rm ln} h}\ ,
  \eeq
which together with the $\beta$-functions of the other couplings
forms a set of coupled differential equations. At one-loop level the
dominant contribution would read $\beta_{\lambda}^{(1)} =
12\lambda^2+6y_t^2\lambda-3 y_{t}^4$, where $y_t$ is the top quark
Yukawa coupling. At higher orders one also has to account for the
coupling to gluons.  To solve for the coupling $\lambda(h)$, we will
employ the next-to-next to leading order code available at
\cite{SMrunning}, which is based on \cite{Chetyrkin:2012rz,
Bezrukov:2012sa}. For the best fit values of the electroweak scale
Higgs mass $m_h=125.7$ GeV,  top mass $m_t=173.1$ GeV and the strong
coupling constant $\alpha_s=0.1184$, the Higgs potential takes the
form shown in Figure 1.
\begin{figure}[!h]
\label{fig:Higgspotdemo}
\begin{center}
\includegraphics[height = 7 cm ]{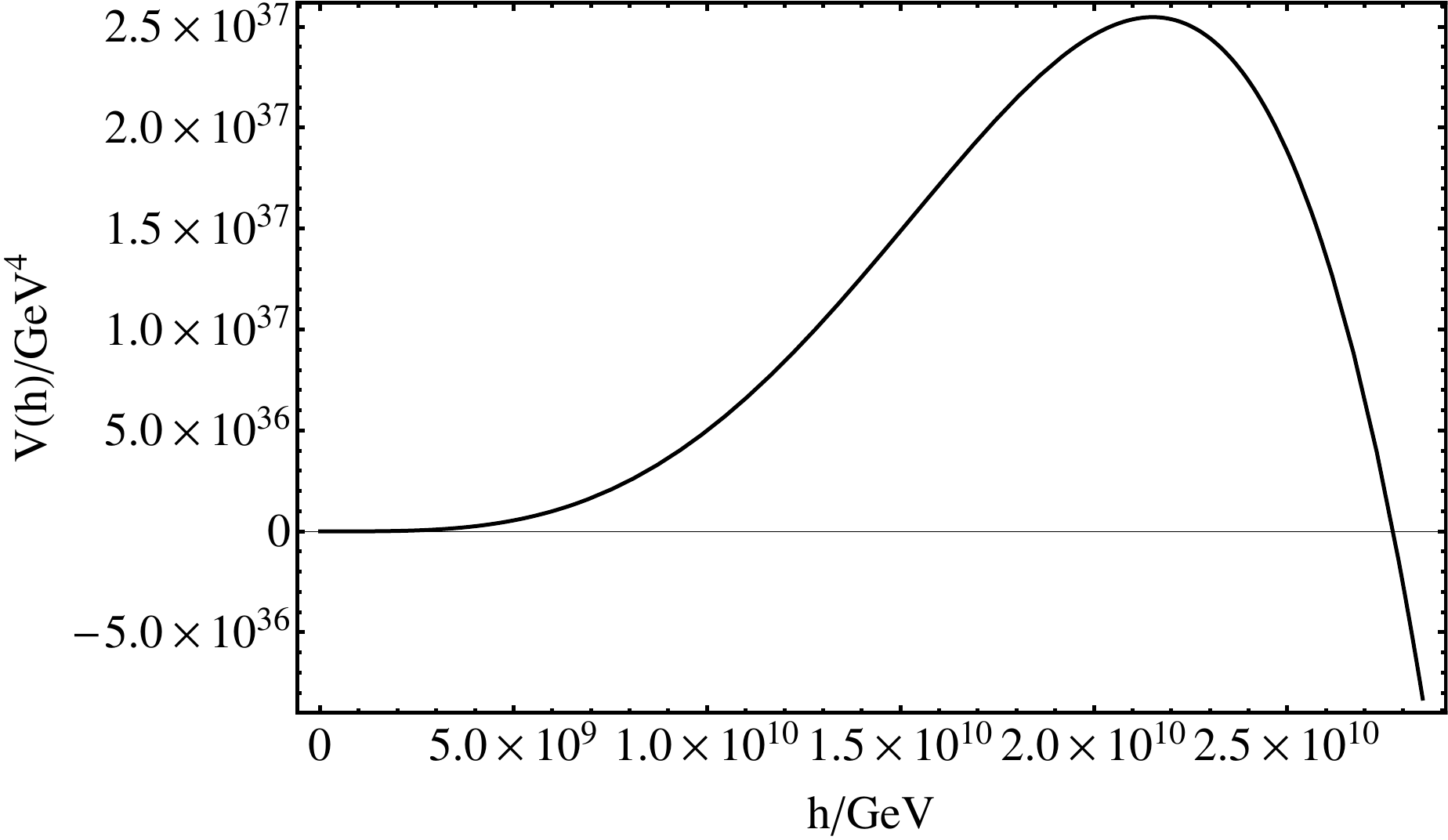}
\end{center}
\caption{The effective potential of the Higgs field, $V=\frac{1}{4}\lambda(h) h^4$, for the best fit SM parameters $m_t=173.1 \, {\rm GeV}$, $\alpha_S=0.1184$ and $m_h=125.7 \, {\rm GeV}$.
Here $h_{\rm max} \simeq 2.1 \times 10^{10}$ GeV and $h_c \simeq 2.8 \times 10^{10}$ GeV.
}
\end{figure}
The local maximum $V'(h_{\rm max})=0$ occurs at
  \beq
  \label{hmax}
  \lambda(h_{\rm max})+\frac{\beta(h_{\rm max})}{4}=0\ .
  \eeq
The maximum eventually vanishes if the Higgs mass is sufficiently
increased or the top mass decreased. Similarly, increasing the Higgs
mass moves the instability scale towards higher field values while
increasing the top mass works in the opposite direction.

Above the instability scale $h>h_{c}$ the SM vacuum would no longer
be the global minimum and new physics must be evoked to restore its
stability. Although not required by theoretical consistency, new
physics could of course appear already at much lower energies. Here
we concentrate on the Higgs dynamics within the SM in the stable
regime $h<h_c$ assuming the new physics does not significantly
affect the potential in this regime.

\subsection{Conditions for consistency}

Without a large non-minimal coupling to gravity, $\xi h^2R$ with
$\xi \gg 1$, the SM Higgs potential (\ref{Vh}) in general is not
flat enough to support slow roll inflation with at least the
required $N_{\rm CMB}\sim 60$ e-folds. For a very fine-tuned choice
of the SM parameters the potential develops a saddle point or a
shallow false minimum. In the vicinity of this point Higgs inflation
could occur for more moderate values of the non-minimal coupling and
even yield the measured tensor-to-scalar ratio
\cite{Hamada:2014iga,Masina:2014yga}.

In the pure SM case the Higgs potential is however too steep to
support inflation, and the same holds true even if small
modifications, such as a small non-minimal coupling $\xi \lesssim
1$, are added to the model. Therefore, if the SM Higgs potential is
not strongly modified at the inflationary scale, inflation should be
driven by new physics. The Higgs energy density must then be
subdominant in order not to spoil the inflationary epoch. This
implies that the allowed range of Higgs values and SM parameters is
constrained by
  \beq
  \label{subdominance}
  V^{1/4}_{\rm SM}(h)\ll \left(3M^{2}_PH^2\right)^{1/4} \simeq 1.6 \times 10^{16} {\rm GeV}\
  .
  \eeq
Here we have used the BICEP2 detection of tensor-to-scalar ratio
$r\simeq0.2$ \cite{Ade:2014xna} to fix the inflationary scale
$\rho_{\rm inf}^{1/4}\simeq 1.6\times 10^{16}$ GeV.

From whatever value the Higgs field initially takes within the
allowed range, it should relax close to the SM vacuum
$h=\nu\simeq246$ GeV well before the electroweak symmetry breaking
crossover. If the Higgs field finds itself beyond the local maximum
$h>h_{\rm max}$ (\ref{hmax}) at some point during inflation, it
should tunnel to the regime $h<h_{\rm max}$ and stay there. As we
will show in the next section, the SM Higgs is a light field for
$h<h_{\rm max}$ and subject to fluctuations generated by the
inflationary expansion. If the energy density stored in the
fluctuations is higher than the height of the Higgs potential at
$h_{\rm max}$, the fluctuations will generically push the Higgs back
to the regime $h>h_{\rm max}$, which is incompatible with the
observed universe.  Requiring the SM vacuum to be stable against inflationary 
fluctuations then yields another constraint between SM parameters and the inflationary scale 
\cite{riotto,higgsinstability}. Taking into account that the kinetic energy of
the fluctuations is $\rho_{\rm kin}(h)\sim H^4$ and using the inflationary 
scale $H\simeq 10^{14}$ GeV implied by BICEP2 the stability condition against 
inflationary fluctuations is given by 
  \beq
  \label{Vmaxbound}
  V^{1/4}(h_{\rm max}) \gtrsim 10^{14} {\rm GeV}\ .
  \eeq
It is readily seen that the Higgs energy density is negligible
(\ref{subdominance}) whenever the above inequality holds. The
condition (\ref{Vmaxbound}) places a direct constraint on the SM
parameters, $m_h$, $m_t$ and $\alpha_s$, which determine the scale
$V(h_{\rm max})$. Therefore, if no new physics modifies the SM up to
the inflationary scale $\rho_{\rm inf}\sim 10^{16}$ GeV, the SM
parameters should lie within the regime depicted in Figure 2
\begin{figure}[!h]
\label{fig:stability}
\begin{center}
\includegraphics[height = 7 cm ]{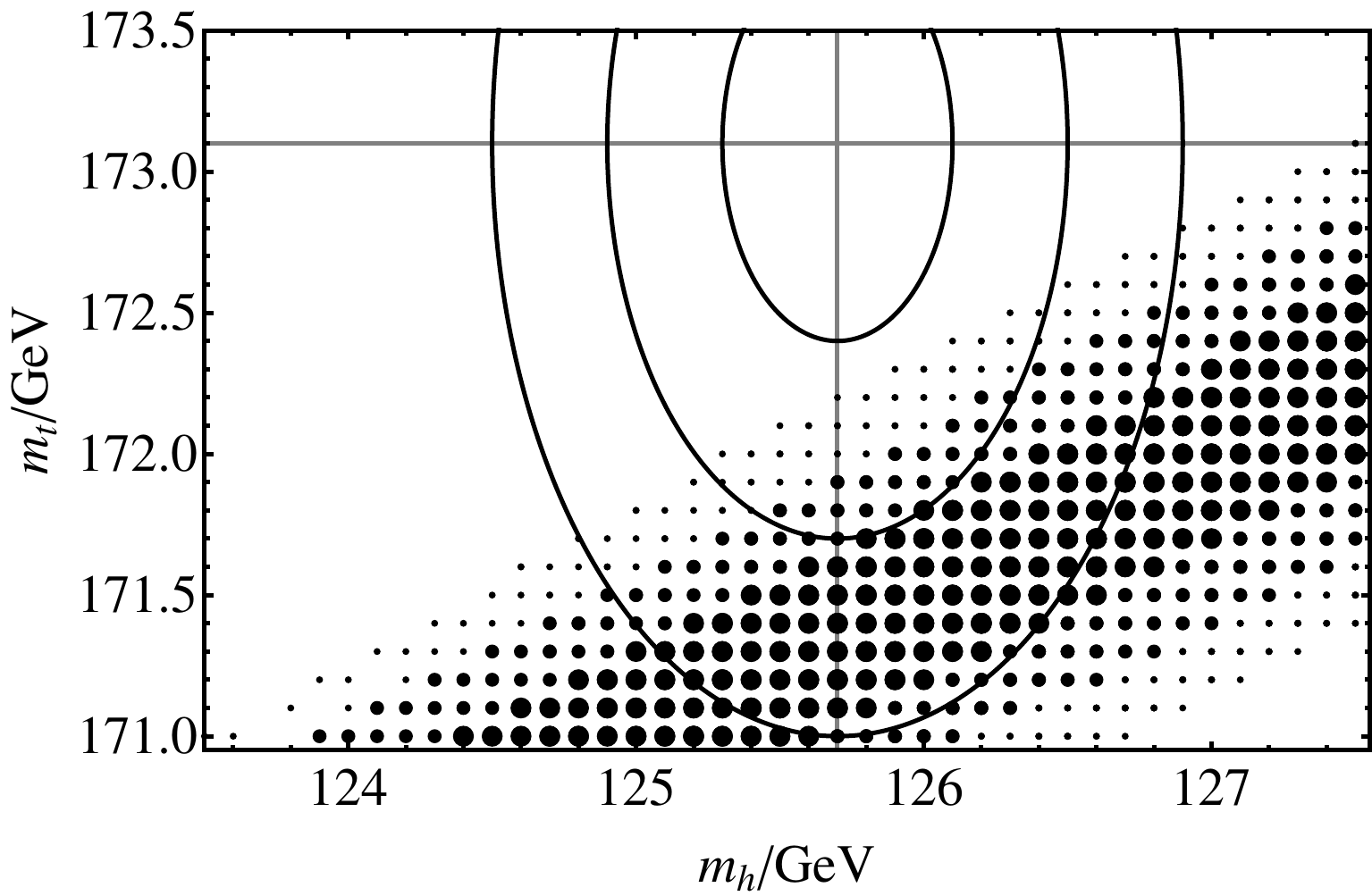}
\end{center}
\caption{The dots depict the top mass $m_{t}$ and Higgs mass $m_h$
region consistent with the observed SM vacuum, pure SM during
inflation and the inflationary scale $H=1.2 \times 10^{14}$ GeV as
implied by the BICEP2 detection. We have marginalized over the
strong coupling constant $\alpha_s=0.1184 \pm 0.0007$ \cite{alphas}, the results marginalized over
the observational $1\sigma$, $2\sigma$ and $3\sigma$ regimes are
depicted respectively by large, medium and small dots. The contours
in the figure depict the $1\sigma$, $2\sigma$ and $3\sigma$ regions
of $m_t=(173.1 \pm 0.7){\rm GeV}$ \cite{Degrassi:2012ry, mtop} and $m_h=(125.7 \pm 0.4){\rm GeV}$ \cite{LHC, mhiggs}. }
\end{figure}
where the inequality (\ref{Vmaxbound}) is satisfied. For parameter
values outside this regime, in particular for the best fit SM
parameter values, the inflationary fluctuations would rapidly push
the Higgs field into the false vacuum $h>h_{\rm max}$, suggesting
that new physics is required to modify the Higgs potential and make
it stable against inflationary fluctuations. The constraint
(\ref{Vmaxbound}) can be satisfied within the SM only if the top
mass is significantly below the best fit value. From the figure, one
finds that the region still consistent with inflation and within
$2\sigma$ errors of the measured SM parameters is given by
\beq
  \label{consistent}
  m_t \lesssim 171.8\,{\rm
  GeV} + 0.538(m_h-125.5\,{\rm GeV} )\, \qquad 125.4\,{\rm GeV}\lesssim
m_h\lesssim 126.3\,{\rm GeV}\ .
  \eeq

One might ask how the constraint (\ref{Vmaxbound}) changes if the
Standard Model is slightly modified. After all, new physics is in
any case needed above the instability scale of the SM and we have
also explicitly assumed that inflation is driven by fields beyond SM
which inevitably couple to SM at least through gravitational
interactions. As long as the new fields coupled to Higgs can be
integrated out during inflation, the changes of the Higgs potential
at the inflationary scale can be encoded into a change in its
coupling $\lambda\rightarrow\lambda+\delta\lambda$. For a modified
SM we can then schematically write a stability condition against
inflationary fluctuations analogous to (\ref{Vmaxbound})
  \beq
  \label{SMmodstability}
  \frac{V_{{\rm SM}+{\rm mod}}(h_{\rm max})}{H^4}= \frac{V_{\rm SM}(h_{\rm
  max})}{H^4}\left(1+\frac{\delta\lambda}{\lambda}\right)\gtrsim 1\ .
  \eeq
Therefore, we find that the condition (\ref{Vmaxbound}) is not
significantly affected unless the modifications are sizeable
$|\delta\lambda|\gg \lambda$.

Let us also note in passing that the new physics might be such that
thermal effects after the end of inflation would lift the false
vacuum. In this case it would be possible to reach the SM vacuum
even if inflationary fluctuations were to push the Higgs field over
the local maximum into the regime $h>h_{\rm max}$. Making more
quantitative statements about such a case would however require the
specification of the unknown nature of new physics. In what follows
we will therefore stick to the condition (\ref{Vmaxbound}) that is
valid within the SM.

%%%%%%%%%%%%%%%%%%%%%%%%%%%%%%%%%%%%%%%%%%%%%%%%%%%%%%%%%%%%%%%%%%%%%%%%%%%
\section{Generation and dynamics of a Higgs
condensate}\label{sec:dynamics}
%%%%%%%%%%%%%%%%%%%%%%%%%%%%%%%%%%%%%%%%%%%%%%%%%%%%%%%%%%%%%%%%%%%%%%%%%%%

Having specified the very generic condition (\ref{Vmaxbound}) for
the consistency of the SM Higgs with the measured inflationary scale
$\rho_{\rm inf}^{1/4}\sim 10^{16}$ GeV, we now move on to study the
Higgs dynamics during inflation in more detail.

As discussed above, if the Higgs finds itself in a false vacuum in
the regime $h>h_{\rm max}$ at some point during inflation, it should
tunnel to the regime $h<h_{\rm max}$ before the end of inflation.
Here we assume the SM potential beyond the instability scale
$h>h_{c}>h_{\rm max}$ is stabilized by some new physics which does
not affect the potential below $h_{\rm max}$. The tunneling
probability is maximized for a process that leaves the Higgs field
at $h=h_{\rm max}$ with a zero kinetic energy. Denoting the
difference between the false vacuum and the local maximum as $\Delta
V=V(h_{\rm max})-V(h_{\rm false})$ (we assume $V(h_{\rm false}) >0$)
the tunneling rate can be estimated by (see e.g. \cite{Linde})
  \beq
  \Gamma/H \sim {\rm exp}\left(-\frac{8\pi^2 \Delta V}{3H^4}
  \right)\ .
  \eeq
Here we have neglected the change of the Hubble scale $H$ as the
Higgs contribution to the total energy density is required to be
negligible (\ref{subdominance}). While the tunneling rate is
suppressed by the condition (\ref{Vmaxbound}), unless the false
vacuum would be very shallow, it could be compensated by a very long
period of inflation before the onset of the observable e-folds. So
if inflation lasts sufficiently long the Higgs could initially start
from the false vacuum. Tunneling could of course take place also
during the observable e-foldings but the probability for this
process is suppressed by the limited number of available of e-folds,
$N_{\rm CMB}\sim 60$.

\subsection{Dynamics close to the local maximum}

Using the next-to-next to leading order result for the radiatively
corrected SM Higgs potential, the effective Higgs mass in the regime
$h < h_{\rm max}$ can be computed. We find that the Higgs is always
massless at the local maximum $m_{h}\ll H^2$ in the regime
consistent with the inflationary scale $H\sim 10^{14}$ GeV implied
by BICEP2 and less than $2\sigma$ away from the measured SM
parameters. Allowing for deviations at $3\sigma$ level, we find that
the Higgs could be massive at $h_{\rm max}$ but even in this case it
becomes massless within $N\lesssim 5$ e-folds. For all practical
purposes we can therefore treat the SM Higgs as a light field after
it has tunneled away from the false vacuum.

Immediately after the tunneling to $h_{\rm max}$ the classical force
vanishes as $V'(h_{\rm max})\sim 0$, and the light Higgs field then
undergoes random walk in the vicinity of $h_{\rm max}$. As the
different regions of the $h = h_{\rm max}$ bubble become stretched
out of causal connection by the inflationary expansion the
stochastic Higgs evolution away from $h_{\rm max}$ in general
differs from patch to patch. In each patch the stochastic epoch
eventually ends when the field has drifted to the point where the
classical force $V'=-3H\dot{h}$ equals the quantum source term
$\delta h/\delta t \sim H^2/2\pi$
  \beq
  \label{hcl}
  V'(h_{\rm cl}) = -\frac{3H^3}{2\pi}\ .
  \eeq
We only concentrate on the patches where $h_{\rm cl}<h_{\rm max}$
and the classical drift in the regime $h<h_{\rm cl}$ drives the
Higgs field towards the SM vacuum. The other patches where $h>h_{\rm
max}$ will relax back to the false vacuum and cannot describe the
observable universe unless the Higgs again tunnels to $h_{\rm max}$
and ends up rolling away from the false vacuum.

While the duration of the stochastic epoch differs from patch to
patch we may estimate its typical time scale by investigating the
behaviour of the two-point function $\langle h^2 \rangle$. The
probability distribution $P(h,t)$ for the Higgs field on
superhorizon scales obeys the Fokker-Planck equation
\cite{starobinsky&yokoyama} which also yields the equation of motion
for the variance $\langle h^2\rangle =\int \d h \,h^2 P(h,t)$
\cite{Tsamis:2005hd}
  \beq
  \label{fokker-planck_variance}
  \frac{\partial \langle h^2\rangle}{\partial t} =
  \frac{H^3}{4\pi^2}-
  \frac{2}{3H}\langle h V'(h)\rangle\ .
  \eeq
The first term on the right hand side represents the contribution of
quantum fluctuations, and the second term corresponds to the
classical drift which starts to grow as the field moves away from
the local maximum $V'(h_{\rm max})=0$. Expanding the potential up to
second order in the displacement $h-h_{\rm max}$
  \beq
  \label{V_expand}
  V(h)=V(h_{\rm max})\left(1 + \frac{3}{2}H^3\eta_{\rm max}(h-h_{\rm max})^2\right)\
  ,
  \eeq
we can solve equation (\ref{fokker-planck_variance}) for the
variance as
  \beq
  \label{variance_sol}
  \left\langle (h-h_{\rm max})^2 \right\rangle= \frac{H^2}{8\pi^2\eta_{\rm max}}\left(1-{\rm exp}\left(-2\eta_{\rm max}
  N\right)\right)\ ,
  \eeq
We solve for the number of e-folds by equating the root mean square of the
variance to the limiting field value of the classical regime
(\ref{hcl}) $\sqrt{\langle(h-h_{\rm max})^2\rangle}=h-h_{\rm cl}$. Thus we
find an estimate for the typical duration of the stochastic epoch
after the tunneling as
  \beq
  \label{eq:Neq}
  N_{\rm cl}=\frac{{\rm ln}\,2}{2 \mid \eta_{\rm max} \mid} \ .
  \eeq
For the SM parameters consistent with the vacuum stability against
inflationary fluctuations, depicted in Figure 2, we find $N_{\rm
cl}\lesssim 20$.

The actual time when the classical regime $h_{\rm cl}$ is reached
differs from patch to patch and is fluctuating around the average
$N_{\rm cl}$. If the currently observable scales exited the horizon
well after this epoch the differences are unobservable as the
subsequent classical evolution carries no memory of the stochastic
epoch. On the other hand, if the observable scales were still inside
the horizon when the field value $h_{\rm cl}$ was reached, the
slight differences in the expansion history might generate
non-trivial features in the spectrum of Higgs perturbations. If the
fluctuations of the subdominant Higgs condensate are converted to
primordial perturbations after the end of inflation this structure
could be imprinted in the CMB perturbations.

\subsection{Intermediate stage and asymptotic dynamics}

The dynamics of the Higgs field in each Hubble patch becomes
dominated by the classical drift $3H{\dot h}\simeq -V'$ when
the field has rolled down to $h_{\rm cl}$ (\ref{hcl}). As the field
keeps rolling towards the minimum, the slope $V'(h)$ starts to
decrease and eventually the dynamics again becomes dominated by the
stochastic quantum noise. This happens for $h<h_{\rm as}$, where
  \beq
  V'(h_{\rm as}) = -\frac{3H^3}{2\pi}\ ,\qquad h_{\rm as} <  h_{\rm
  cl}\ .
  \eeq
For the SM parameter values in Figure 2 consistent with $H_{\rm
inf}\sim 10^{14} $ GeV we find that the Higgs can stay at most
$N_{\rm int}\lesssim 70$ e-folds in the classical regime $h_{\rm
as}<h<h_{\rm cl}$. As the field has rolled down to $h_{\rm as}$ its
motion becomes dominated by the quantum fluctuations. The transition
from the classical to stochastic epoch could again leave observable
imprints into the primordial perturbations sourced by the Higgs
condensate provided the transition takes place when observable
scales are leaving the inflationary horizon.

In the asymptotic stochastic regime $h< h_{\rm as}$ the classical
drift towards the Standard Model vacuum gets overwhelmed by the
backreaction of the generated quantum fluctuations and field starts
to undergo a random walk. At the onset of the stochastic epoch the
probability distribution of the Higgs field over a horizon patch
is peaked around $h_{\rm as}$ and then starts to spread out and move
towards the equilibrium distribution \cite{starobinsky&yokoyama}
  \beq\label{equlibrdist}
  P(h)\simeq C\, {\rm exp}\,\left(-\frac{8\pi^2V(h)}{3
  H^4}\right)\ .
  \eeq
The stability condition (\ref{Vmaxbound}) guarantees that the
equilibrium probability to fluctuate to the regime of the false
vacuum is heavily suppressed. We can then normalize the
probability within the regime $|h|<|h_{{\rm max}}|$ so that
  \beq
  C^{-1}=\int_{-h_{\rm max}}^{h_{\rm max}}dh\: P(h)\ .
  \eeq

In the asymptotic regime $h <  h_{\rm as}$ the running of the
coupling $\lambda(h)$ in (\ref{Vh}) is a small effect and the Higgs
potential is nearly quartic with $V\sim h^4$. For a quartic
potential, the spreading of an initial probability distribution
towards the equilibrium result (\ref{equlibrdist}) is characterized
by a decoherence time, which in terms of e-folds has been found
\cite{Enqvist:2012xn} to be given by $N_{\rm dec}\approx
6\lambda^{-1/2}$. Using this estimate for the SM Higgs we find
$N_{\rm dec}\lesssim 100$. Therefore, if there was at least $N_{\rm
cl}+N_{\rm int}+N_{\rm dec} = {\cal O}(200)$ e-folds of inflation
from the tunneling of the Higgs to $h_{\rm max}$ until the horizon
exit of the observable scales, the Higgs amplitude is controlled by
the equilibrium distribution (\ref{equlibrdist}) at the time during
which the observable CMB scales are leaving the horizon. An estimate
of the typical Higgs value in our patch is then given by the root
mean square $h = \sqrt{\langle h^2\rangle}$ computed from
(\ref{equlibrdist}). Treating the Higgs coupling $\lambda(h)$ as a
constant one then finds \cite{Enqvist:2013kaa}
  \beq
  \label{Higgsvalue}
  h\simeq 0.4 \lambda^{-1/4}
  H\sim 10^{14} {\rm GeV}\
  \eeq
for the inflationary scale $H\sim 10^{14}$ GeV implied by BICEP2
\cite{Ade:2014xna}. We have checked that including the running of
the coupling $\lambda(h)$ in (\ref{equlibrdist})  does not
significantly change the result within the parameter range
consistent with the stability condition (\ref{Vmaxbound}) as
depicted in Figure 2.

%%%%%%%%%%%%%%%%%%%%%%%%%%%%%%%%%%%%%%%%%%%%%%%%%%%%%%%%%%%%%%%%%%%%%%%%%%%
\section{Discussion}\label{sec:discussion}
%%%%%%%%%%%%%%%%%%%%%%%%%%%%%%%%%%%%%%%%%%%%%%%%%%%%%%%%%%%%%%%%%%%%%%%%%%%

We have considered the constraints imposed on the Standard Model by
the assumption that up to the inflationary scale, the Higgs
potential is at least approximatively given by the pure SM
prediction and not significantly affected by the field(s) driving
inflation. These constraints are of cosmological nature and follow
from the fact that during inflation, for all practical purposes the
SM Higgs is a light field, which we have verified. Thus during
inflation the Higgs field is subject to fluctuations: there will be
local field perturbations, but in addition, also the mean field
performs a random walk. If inflation lasts long enough, about 200
efolds, the mean field will have settled into its equilibrium
distribution, that can be derived in the stochastic approach, by the
horizon exit of observable scales. This will provide the initial
condition for the Higgs condensate after inflation which is an
integral part of the initial data for the subsequent hot big bang
epoch.

For the best fit parameters and in the next-to-next leading order,
the potential of the SM Higgs has a local maximum at large field
values,  $h_{\rm max}\sim 10^{10}$ GeV. Beyond the maximum there is
a false vacuum, which can be either stable or unstable. If unstable,
it should be stabilized by new physics modifying the SM potential
above scale of the local maximum. The basic assumption here is that
new physics has no significant impact on the Higgs potential at
field values below  $h_{\rm max}$.

Whatever the value the Higgs field had at the end of inflation, it
should relax to the SM vacuum by the time the electroweak symmetry
breaking takes place. Unless the false vacuum gets lifted by thermal
corrections after inflation, or is extremely shallow, this
requirement implies that the Higgs field at the end of inflation
must be at or below the local maximum $h_{\rm max}$ so that it can
relax into the correct vacuum by classical dynamics. 
Here we have pointed out, see also \cite{riotto,higgsinstability}, 
that for the best fit values this requirement is in
tension with the high inflationary scale inflationary scale $H_{\rm
inf}\sim 10^{14}$ GeV implied by the BICEP2 detection of
gravitational waves.

During inflation the SM Higgs turns out to be effectively massless
for field values below the local maximum. Hence the mean field
acquires fluctuations proportional to the inflationary scale $\delta
h\sim H_{\rm inf}\sim 10^{14}$ GeV. Therefore, it is not enough that
during inflation the Higgs is located below $h_{\rm max}$ when the
observable scales exit the horizon.  This configuration has to be
also stable against inflationary fluctuations, which could carry the
mean field over into the false vacuum. We argue that the tunneling
rate out of the false vacuum should be negligible over the
observable e-folds. We then show that the condition for the
stability is given by $V(h_{\rm max}) \gtrsim H^4$, where $V(h_{\rm
max})$ is the potential energy at the local maximum. Computing
$V(h_{\rm max})$ in the next-to-next leading order, we find that the
SM Higgs the stability is guaranteed only for a sufficiently low top
mass with 2-3 $\sigma$ below the best fit value, depending on the
measured values of $m_h$ and $\alpha_s$.

There may be particle physics reasons for extending the Standard
Model, but if the still allowed parameter region depicted in Fig. 2
can be ruled out, the observed high inflationary scale alone would
require new physics modifying the Higgs potential. The required
modifications should be significant as moderate shifts
$|\delta\lambda|\sim \lambda_{\rm SM}$ of the effective Higgs
coupling from its SM value at the inflationary scale would not
affect the orders of magnitude in the stability condition $V(h_{\rm
max}) \gtrsim H^4$. Note that since $H\propto r^{1/2}$ our
conclusion is also not sensitive to the exact value of the tensor-to-scalar
ratio. Even if the observed tensor-to-scalar ratio would go
significantly down from $r\sim 0.2$ the SM vacuum for the best fit
parameter values would remain unstable against inflationary
fluctuations.

We have also carefully investigated the Higgs dynamics during
inflation for the SM parameters consistent with the stability
condition $V(h_{\rm max}) \gtrsim H^4$. We have argued that the
transitions between classical and stochastic regimes in the Higgs
dynamics could leave distinct imprints in the spectrum of Higgs
fluctuations. If the transitions occur when the observable scales
leave the horizon, and if the Higgs perturbations source either
adiabatic or isocurvature metric fluctuations, these imprints could
be observable in the CMB.

While the paper was in preparation, there appeared an article \cite{Fairbairn:2014zia} which also discusses SM stability in the light of BICEP2, with which our results are in a qualitative agreement.

\acknowledgments{KE is supported by the Academy of Finland grants
1263714 and 1263714; TM is supported by the Magnus Ehrnrooth
foundation, and SN is supported by the Academy of Finland grant
257532.}

\end{document}